\documentstyle[aps,prb,preprint]{revtex}

\begin{document}
\draft
%\preprint{incoherent2.tex}
\title{Origin of Intrinsic Josephson Coupling in the Cuprates\\
and Its Relation to Order Parameter Symmetry:\\
An Incoherent Hopping Model}
\author{R. J. Radtke}
\address{Center for Superconductivity Research, Department of Physics,\\
University of Maryland, College Park, Maryland, 20742-4111}
\author{K. Levin}
\address{Department of Physics and The James Franck Institute,\\
The University of Chicago, Chicago, Illinois  60637}
\maketitle

\begin{abstract}
Experiments on the cuprate superconductors demonstrate
that these materials may be viewed as a stack of Josephson junctions
along the direction normal to the CuO$_2$ planes (the $c$-axis).
In this paper, we present a model which describes this intrinsic
Josephson coupling in terms of incoherent quasiparticle hopping
along the $c$-axis arising from wave-function overlap,
impurity-assisted hopping, and boson-assisted hopping.
We use this model to compute the magnitude and temperature $T$
dependence of the resulting Josephson critical current $j_c (T)$ for
$s$- and $d$-wave superconductors.
Contrary to other approaches, $d$-wave pairing in this model is
compatible with an intrinsic Josephson effect at all hole
concentrations and leads to $j_c (T) \propto T$ at low $T$.
By parameterizing our theory with $c$-axis resistivity data from
$\rm YBa_2Cu_3O_{7-\delta}$ (YBCO), we estimate $j_c (T)$ for
optimally doped and underdoped members of this family.
$j_c (T)$ can be measured either directly or indirectly through
microwave penetration depth experiments, and current measurements
on $\rm Bi_2Sr_2CaCu_2O_8$ and $\rm La_{2-x}Sr_xCuO_4$
are found to be consistent with $s$-wave pairing and the dominance
of assisted hopping processes.
The situation in YBCO is still unclear, but our estimates suggest
that further experiments on this compound would be of great help
in elucidating the validity of our model in general and the
pairing symmetry in particular.
\end{abstract}


\section{Introduction}
While much of the early work on the cuprates has focused on the
copper oxide planes, recent experimental\cite{Cooper}
and theoretical\cite{KumarL,Tachiki,Klemm,KumarJ,Leggett,Graf,Rojo} efforts
are beginning to address the $c$-axis properties.
In large part, this interest is motivated both by the
availability of large, high-quality single crystals and by the
realization that a full understanding of the superconductivity cannot be
reached until the role of the third direction is established.
Indeed, the third dimension is ssential for the high critical
temperatures in some theories.\cite{AZ,Chakravarty}

While a complete theoretical description of the $c$-axis properties
of the cuprates is clearly desirable, a full understanding of these
properties is hampered by their unusual experimental signatures.
In the normal state, for example,
the resistivity along the $c$-axis ($\rho_c$) is quite
generally metallic in the optimally doped materials and becomes
semiconducting when the hole doping is reduced, while the
in-plane resistivity ($\rho_{ab}$) always shows metallic
behavior.\cite{Cooper}
At first sight, the metallic behavior of both $\rho_{ab}$ and $\rho_c$
in the optimally doped compounds
is understandable within the conventional Bloch-Boltzmann
theory of transport; however, the semiconducting temperature dependence
of $\rho_c$ with metallic $\rho_{ab}$ in the underdoped samples
is difficult to reconcile with this picture.
In the superconducting state, also, measurements of hysteretic
current-voltage curves, Fraunhofer-like $c$-axis critical currents
as a function of in-plane magnetic field, and emission of
microwave radiation at ac Josephson effect frequencies strongly
indicate that the cuprates can be viewed as a stack of SIS Josephson
junctions along the $c$-axis.\cite{KM}
This situation is not at all consistent with the properties
of conventional superconductors.

One picture which may account for these anomalous
properties is based on incoherent $c$-axis transport.
In conventional metals, the presence of the crystal lattice
leads to the formation of three-dimensional Bloch waves.
The basic postulate of the incoherent approach is that
Bloch waves in the $c$-direction do not form due to the layered
structure, intrinsic
disorder, and/or strong electronic correlations in the cuprates.
This situation is not equivalent to localization of the quasiparticles
in the $c$-direction, though, because they
may still hop {\it in}coherently between the CuO$_2$ layers.
Working under this assumption,
we will develop a theory which explains the unusual
temperature-dependence of $\rho_c$ in the normal state and
the intrinsic Josephson effect in the superconducting state
and which may also shed light on the question of the order parameter
symmetry in these systems.

The specific model we adopt was first articulated by Rojo and
Levin\cite{Rojo} and asserts that the CuO$_2$ layers can be
thought of as independent 2D Fermi liquids weakly coupled along
the $c$-axis by three processes:
direct inter-layer quasiparticle hopping,\cite{KumarL,KumarJ}
hopping assisted through static disorder,\cite{Graf}
and hopping assisted through dynamic, boson-mediated (e.g.,
phonon-mediated) scattering.\cite{KumarL,Rojo}
The semiconducting temperature dependence of $\rho_c$
is then associated with the reduction of the
dynamic inter-planar scattering as the temperature is lowered.
The $c$-axis conduction is therefore analogous to that of
a very dirty, but nevertheless metallic, system.\cite{Rojo}

We will show that all three of these processes give rise
to Josephson coupling, although their relative importance depends
on the material and on the order parameter symmetry.
It should be no surprise that all three mechanisms produce
Josephson coupling between the layers;
direct and impurity-assisted hopping processes formally resemble
the tunneling processes considered in conventional SIS
junctions,\cite{Cohen,AB,Duke,TMahan}
and boson-assisted hopping has been known to contribute to
tunneling in superconducting junctions in both the
quasiparticle\cite{Kleinman} and Josephson\cite{Duke} channels
for some time.
In the cuprates, interest in these mechanisms has been piqued by
the interpretation of experiments involving the $c$-axis
transport\cite{Cooper} and flux quantization in superconducting
rings.\cite{Kirtley,Bulaevskii}
This interest is reflected in a rapidly developing theoretical literature.
In particular, we mention related results on intrinsic Josephson
coupling in the cuprates by Graf {\it et al.},\cite{Graf2} which
came to our attention after completion of this work.
As in this paper, these authors employ an incoherent hopping model
to describe the inter-planar coupling in the cuprates.
We differ in that Graf {\it et al.} consider only
impurity-assisted hopping while we also include direct and
boson-assisted processes.
As a result, we find a non-zero $d$-wave Josephson critical current in
fully oxygenated YBCO without the need to introduce an anisotropic
impurity scattering rate.
In addition, because the present approach can describe the temperature
dependence of the normal-state transport, we have been able to compute
the full temperature dependence of the critical currents.

We present our calculations as follows.
In the next Section, we describe our model and the
approximations used in our calculations.
For later reference, we will also discuss the normal-state $c$-axis
conductivity $\sigma_c$ and derive expressions for the contributions
to this conductivity by the three processes.
We then move on to the superconducting state and derive the general
forms of the Josephson critical current $j_c$ which arise from each
inter-layer hopping mechanism.
The direct and impurity-assisted critical currents are found to
have familiar forms, but the boson-assisted contribution gives rise to
a novel expression.
Finally, we quantify our model by parameterizing experimental
resistivity data in the $\rm YBa_2Cu_3O_{7-\delta}$ (YBCO) system and
using the estimated parameters to predict $j_c(T)$ for different oxygen
stoichiometries and pairing symmetries.
Since $j_c$ can be measured directly\cite{KM} or inferred from microwave
penetration depth experiments,\cite{Clem} our predictions may be
compared to experiment.
Based on these calculations, we conclude that low-temperature measurements
of $j_c (T)$ should be able to distinguish $s$- from $d$-wave
pairing regardless of oxygen stoichiometry.
Moreover, we find that the $c$-axis penetration
depth ratio should have either a 3D BCS character if the pairing
is $d$-wave or a 3D BCS character which crosses over as a function
of hole doping to a 2D Ambegaokar-Baratoff character if the
pairing is $s$-wave.
Although experiments have not converged in the YBCO system, we
discuss the implications of our model for the pairing symmetry
in the other cuprates in the last subsection.
A preliminary report of these results has been published
elsewhere.\cite{Lau}

\section{Incoherent Hopping Model}

\subsection{Description and Assumptions}

We begin by writing the Hamiltonian for the electronic system in terms
of the Hamiltonians of the individual CuO$_2$ layers $H_m$
and an inter-layer coupling Hamiltonian $H_\bot$:
$H_{\rm el} = \sum_m H_m + H_{\bot}$.
Here, the term ``layer'' denotes a composite object
consisting of all the CuO$_2$ planes within a single unit cell,
so the transfer Hamiltonian $H_\bot$ takes
quasiparticles from one unit cell to another.
Motivated by Ref.~\onlinecite{Rojo}, we take
$H_{\bot} = A^{ } + A^{\dag}$, with
\begin{equation}
A =  \sum_{im\sigma} (t_{\bot} + V_{im} + \sum_j g_{i-j,m} \phi_{jm})
c_{i,m+1,\sigma}^{\dag} c_{im\sigma}^{ }
\label{eq:a}
\end{equation}
and $c_{im\sigma}$ a quasiparticle annihilation operator for
site $i$ in layer $m$ and spin projection $\sigma$.
The terms in $H_{\bot}$ represent a direct hopping arising from
wave-function overlap ($t_{\bot}$), an impurity-assisted hopping
due to static disorder of strength $V_{im}$, and a boson-assisted
hopping represented by the field $\phi_{im}$ with coupling strength
$g_{im}$.
The total Hamiltonian is then the electronic Hamiltonian added to
the Hamiltonian governing the dynamics of the $\phi$ field:
$H_{\rm tot} = H_{\rm el} + H_{\phi}$.

In studying the properties of this model, we make two
assumptions;
the first relates to how we include incoherence in the model
and the second to how we view the intra-layer dynamics.
Microscopically, incoherence may arise from strong intra-layer
scattering,\cite{KumarJ} a non-Fermi-liquid ground state within the
layers,\cite{AZ} dynamic inter-layer scattering,\cite{KumarL} or
strong electromagnetic fluctuations.\cite{Leggett}
We do not attempt to construct a fully microscopic theory of this
incoherence, but simulate it by taking $H_{\bot}$ to be
a small perturbation on $H_m$ and calculating its effect within
second order perturbation theory.
This approach is similar in spirit to that of
Ioffe {\it et al.}\cite{Ioffe} and should yield the correct
features of a more complete theory.
In order to perform quantitative calculations within this model,
we also assume
that the normal state of $H_m$ for each layer
$m$ is a Fermi liquid and that the superconducting state may be described
within Migdal-Eliashberg theory.\cite{AM}

{}From these two assumptions, the procedures outlined in the rest
of this paper may be applied to any type of inter-layer disorder
$V_{im}$ or inelastic scattering $\phi_{im}$.
For concreteness, however, we specify these parameters.
First, we take the disorder to be delta-function
correlated:  $\overline{V_{im}} = 0$ and
$\overline{V_{im} V_{jn}} = V^2 \delta_{ij} \delta_{mn}$, where
the overbar denotes the usual average over impurities.\cite{AGD}
Moreover, we take the electron-inter-layer-boson coupling to be
structureless
($g_{im} = g$) and the bosons themselves to be represented by
an Einstein spectrum of frequency $\Omega_0$.
These choices do not affect the character of our results in
$s$-wave superconductors, but they do affect the critical currents
in $d$-wave superconductors in a way which will be discussed in
Section~III.

When it is necessary to include intra-layer scattering processes
in our calculations, we do so through a temperature-dependent
scattering rate $\hbar / \tau_{ab} (T)$ which is linear
in $T$ at optimal doping, at least down to the critical temperature $T_c$.
In addition, all wave-vector integrals are performed by the usual
technique of restricting the wave vectors to the Fermi surface,
taking the density of states to be constant, and then
integrating over the remaining energy dependence.
This approximation is appropriate if the Fermi level is far from
any structure in the density of states (e.g., Van Hove singularities)
and the other energy scales in the problem are much less than the
band width.
We assume these conditions hold here.

\subsection{Normal-State Conductivity}

To illustrate the application of our model and for future
reference, we compute in
this subsection the normal-state $c$-axis conductivity within the
incoherent hopping model outlined above.
By assumption, this calculation is to be carried out to second
order in $H_\bot$ and thus becomes formally identical to computing
the conductivity of an NIN tunnel junction, with $H_\bot$ playing the
role of the transfer Hamiltonian.\cite{AZ,LV}
Applying the standard theory of tunneling,\cite{TMahan}
we compute the correlation function
$X (\tau) = -\left< T_{\tau} \left[ A^{ } (\tau) A^{\dag} (0) \right] \right>$
with $A$ defined by Eq.~(\ref{eq:a}) and with the $\tau$-dependence
of $A$ determined by $\sum_m H_m + H_{\phi}$.
The diagrams used in this calculation are shown in
Fig.~\ref{fig:diagrams};
note that vertex corrections to $X (\tau)$ are of higher order
in the inter-layer hopping amplitudes and so are neglected.
Analytically continuing the Fourier transform of
$X (\tau)$ to real frequencies
gives the $c$-axis current density produced by a voltage $V$ across
nearest neighbor layers as
$j_c(V) = - (2e / \hbar a^2) \, {\rm Im} X (eV)$, from which the
conductivity $\sigma_c = c \, (\partial j_c(V) / \partial V)|_{V = 0}$
is obtained.
(Throughout this paper, $a$ ($c$) denotes the intra- (inter-) layer
unit cell dimension).
The resulting conductivity is the sum of three terms, each
corresponding to one of the inter-layer hopping processes:
$\sigma_c = \sigma_c^{\rm direct} + \sigma_c^{\rm imp} + \sigma_c^{\rm inel}$.

The first term in $\sigma_c$ results from the direct hopping
characterized by $t_{\bot}$ and can be written
\begin{equation}
\sigma_c^{\rm direct} = \sigma_0 \, N_0 t_{\bot}^2 \,
  \left( \frac{\tau_{ab}}{\pi \hbar} \right),
\label{eq:o_direct}
\end{equation}
where $N_0$ is the density of states per unit cell
per spin at the Fermi surface and
$\sigma_0 = (4 \pi e^2 / \hbar) (c / a^2)$,
Up to factors of order unity, this conductivity reproduces the results
of Kumar and Jayannavar\cite{KumarJ} and Ioffe {\it et al.}\cite{Ioffe}
Note that $\sigma_c^{\rm direct}$ is proportional to the intra-layer
lifetime $\tau_{ab}$, just as in the usual Drude expression for the
conductivity,
despite the fact that the conductivity is viewed as a tunneling
process.
To understand this surprising result, recall that the Kubo
formula for the conductivity in 3D metals leads to a particle-hole
bubble with group velocity factors at the vertices, while these
vertices become the tunneling matrix element $T_{\bf k^\prime k}$
for tunneling calculations.
The Kubo conductivity thus has a different form than
the tunneling conductivity in 3D materials.
For the 2D metals comprising the layers in our model, however,
the tunneling matrix element for direct hopping is
$T_{\bf k^\prime k} = t_\bot \delta_{\bf k^\prime k}$,
and so vertices are diagonal in 2D wave vector.
Consequently, the Kubo and tunneling formalisms give the same
results.\cite{PhD}

The second and third terms in $\sigma_c$ result from the assisted
processes and are found to be
\begin{equation}
\sigma_c^{\rm imp} = \sigma_0 \, (N_0 V)^2
\label{eq:o_imp}
\end{equation}
for impurity-assisted hopping and
\begin{equation}
\sigma_c^{\rm inel} = \sigma_0 \, (N_0 g)^2
  \frac{\hbar \Omega_0 / 2 k_B T}{\sinh^2 \, (\hbar \Omega_0 / 2 k_B T)}.
\label{eq:o_ba}
\end{equation}
for boson-assisted hopping.
Observe that the impurity-assisted hopping conductivity is
the usual NIN tunneling conductivity.\cite{TMahan}
Also note that the procedure in Ref.~\onlinecite{Rojo} of simulating
the presence of boson-assisted hopping in $\sigma_c$ by allowing the
impurity-assisted
hopping amplitude to become a function of temperature is microscopically
justified by the more complete calculations which yield
Eqs.~(\ref{eq:o_imp})-(\ref{eq:o_ba}).

The form of the conductivity given by Eqs.~(\ref{eq:o_direct})-(\ref{eq:o_ba})
provides a simple explanation for the
anomalous normal-state resistivity in the cuprates.
At optimal stoichiometry, we expect the layers to be strongly coupled
and nearly coherent, implying that the direct hopping should dominate
$\sigma_c$.
Consequently, $\rho_c \propto \rho_{ab}$ as in conventional
Bloch-Boltzmann transport.\cite{BB}
Off optimal stoichiometry, the direct inter-layer coupling may weaken
relative to the assisted processes, making the material increasingly
incoherent.
While this change would not strongly affect $\rho_{ab}$, $\rho_c$
would now be dominated by the assisted hopping, the inelastic component
of which would freeze out at low $T$ [cf. Eq. (\ref{eq:o_ba})].
Thus, as $T$ decreases, the net inter-layer hopping amplitude would be
reduced and $\rho_c$ would increase, as observed experimentally.
These qualitative statements can be made quantitative through
detailed fits of the normal-state resistivity to the Rojo-Levin
model.\cite{Lau,PhD}

\section{Josephson Critical Current}

The incoherent hopping model described above views the inter-layer
transport as an NIN tunneling process.
In the superconducting state, this formulation of the transport
immediately produces an intrinsic Josephson effect between the layers.
In this Section, we compute the temperature-dependent critical
current associated with each inter-layer hopping process
to determine the general features of the total $j_c (T)$.
In the next Section, we will apply these theoretical results to
the cuprates and see what can be revealed about the order parameter
symmetry and the relative importance of the inter-layer hopping
mechanisms.

\subsection{Formalism}

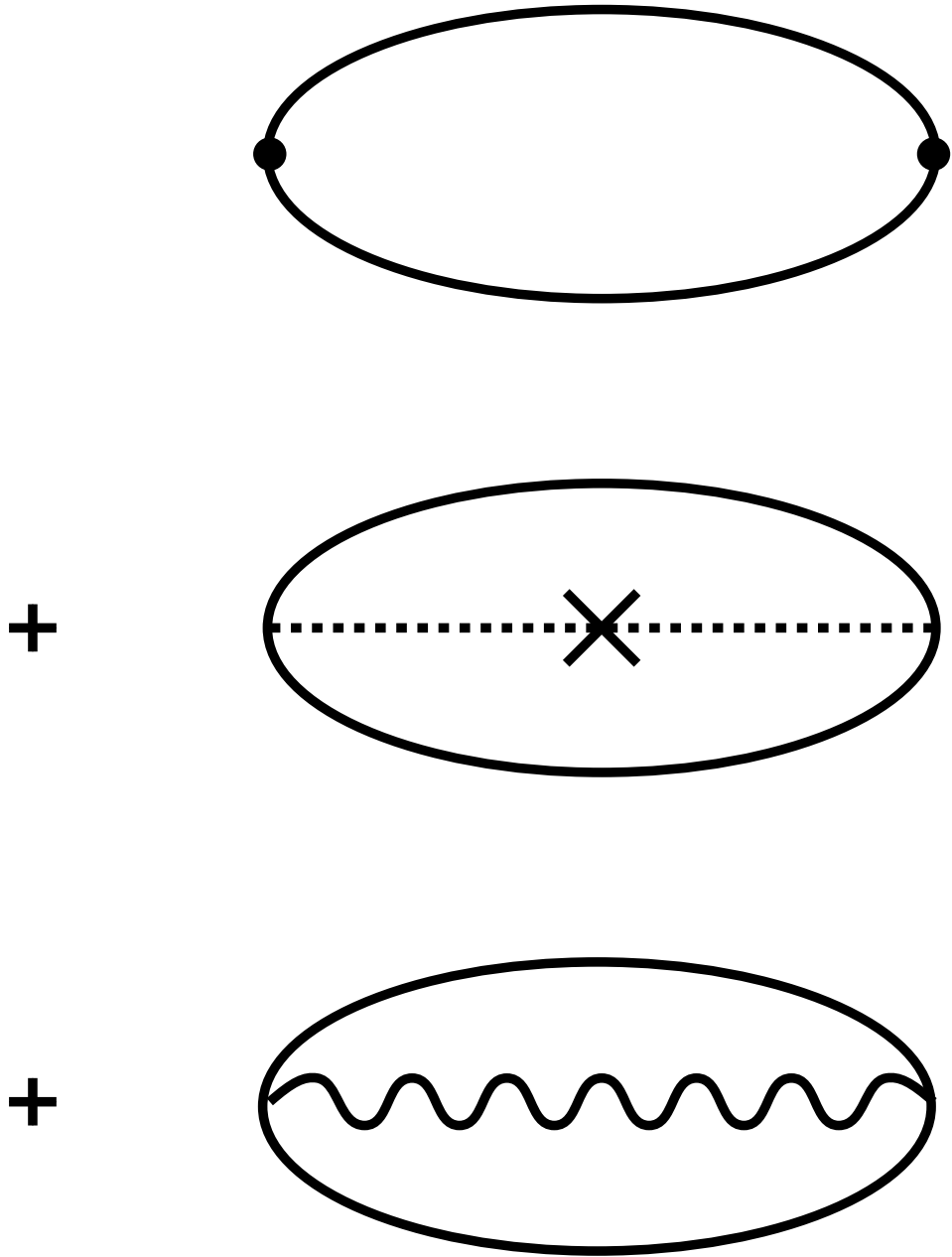
\begin{figure}
\caption{Feynman diagrams used to compute the $c$-axis conductivity
and Josephson critical current within the incoherent hopping model.
The solid dots at the vertices represent the direct inter-layer
hopping amplitude $t_\bot$, the dashed line with a cross represents
the averaged inter-layer disorder, and the wavy line represents the
propagator for boson-assisted hopping.
The two electronic Green's functions in each diagram correspond to
nearest-neighbor layers; normal-state propagators are used to
compute the conductivity while Gor'kov propagators are used in
the superconducting state.
See text for details.}
\label{fig:diagrams}
\end{figure}

The technique for calculating the Josephson critical current
in an SIS junction is well known and is closely related to
the corresponding NIN calculation.
One simply uses the same diagrams from the NIN computation
and replaces the normal Green's functions with anomalous
or Gor'kov Green's functions.\cite{AB,Duke,TMahan}
Specifically, we define the correlation function
$\Phi(\tau) = - \left< T_{\tau} \left[ A (\tau) A (0) \right] \right>$
and evaluate it using the diagrams in Fig.~\ref{fig:diagrams}.
The Fourier transform of the
resulting expression is then analytically continued to real
frequencies to obtain the Josephson current density
produced by an applied voltage $V$:
$j_J (t) = - (2e / \hbar a^2) \, {\rm Im} \,
   \left[ e^{-2ieVt / \hbar} \, \Phi (eV) \right]$.
Setting $V$ = 0 and adjusting the relative phase of the order
parameters on different layers to maximize $j_J (0)$, we obtain
the Josephson critical current as the sum of three components:
\begin{equation}
j_c (T) = j_c^{\rm direct} (T) + j_c^{\rm imp} (T) + j_c^{\rm inel} (T).
\label{eq:j}
\end{equation}
In principle, these critical currents depend not only on the
inter-layer hopping mechanism but also on the order parameter
symmetry.\cite{fluc}

Although the Josephson critical current and the conductivity
are computed from the same set of diagrams, there are significant
differences.
The most obvious is that the propagators used in computing the
diagrams are different; specifically, the phases of the order
parameters enter into the Josephson calculation in a fundamental
way and affect how the imaginary part of the correlation function
$\Phi$ is taken.
Moreover, the conductivity is the derivative of the correlation
function $X$ while the Josephson current is the correlation function
$\Phi$ itself.
We therefore cannot expect similar behavior from these two quantities.
In particular, we will see that the boson-assisted hopping does
not contribute to the conductivity at low $T$ but {\it does} contribute
to the Josephson current.

In evaluating the correlation function $\Phi$, we should ideally
solve the full Eliashberg equations\cite{AM} for the layered system and
use the results of that calculation to construct the anomalous
propagators present in $\Phi$.
This approach would naturally include all the effects of elastic
and inelastic scattering which would naively have a large impact
on the Josephson coupling.
As we will argue below, however, this procedure is not necessary,
since inter-layer scattering effects are of higher order
in the inter-layer hopping amplitudes and the intra-layer
scattering may be incorporated into a BCS-like gap function.
As we shall see, this approach is a natural consequence of our model
and is supported by the long history of experimental and
theoretical work on the Josephson effect in SIS
junctions.\cite{Duke,Bulaevskii}

In performing the reduction from the full Eliashberg equations,
we first note that all quantities in our model are by construction
computed to second order in the inter-layer hopping amplitudes.
This approximation is reasonable
in the superconducting state as long as the hopping amplitudes are
smaller than the maximum of the gap function, as we expect
them to be.
Since the vertices in the diagrams for $\Phi$ are each first order
in the hopping amplitudes [cf. Fig.~\ref{fig:diagrams}],
the complete second-order calculation involves propagators for
only single, isolated layers.
In particular, there are no vertex corrections to $\Phi$ and
any loss of Josephson coupling due to
inelastic inter-layer hopping comes in at higher order in the
inter-layer hopping amplitudes.

Solving for the propagator is facilitated by observing
that the critical current depends only on the weak-coupling propagator
with the strong-coupling gap function $\Delta ({\bf k}, i\omega_n)$
(= $\phi({\bf k}, i\omega_n) / Z({\bf k}, i\omega_n)$ in the
Nambu notation\cite{AM}) in the approximation where
wave vectors are restricted to the Fermi surface and the energy
integrals are extended to infinity.
Hence, all intra-layer elastic and inelastic scattering can be incorporated
into our model by computing their effect on the gap function.
Actually performing this computation for the cuprates
is complicated by the presence of the unusual intra-layer inelastic
scattering, which presumably gives rise to the linear-in-$T$ resistivity.
Since the origin of this scattering is presently controversial,
we are forced to take a phenomenological approach at this point.

Empirically, the cuprates exhibit the generic features of BCS
superconductors, although with quantitative differences in,
for example, $2\Delta / k_BT_c$.
We therefore adopt the conventional picture and compute $j_c$ using
the BCS form for the gap function.
Anderson's theorem,\cite{Anderson} in combination with the
experimental fact that even very dirty tunnel junctions exhibit
Ambegaokar-Baratoff behavior, suggests that this approach
should be reasonable in the $s$-wave case.
For $d$-wave superconductors, the effect of disorder and scattering
is more subtle,\cite{Hirschfeld} so our results should only
be taken qualitatively.

We conclude this subsection by recapitulating some of the
assumptions which underlie the calculations which follow.
First and foremost, we have simulated the effects of incoherent
inter-layer transport through second-order perturbation theory.
This treatment does not address the details of how the
incoherence arises in the first place, so further work will be
required in order to obtain a fully self-consistent description.
We have also neglected the detailed structure of the cuprates within
a unit cell (multiple CuO$_2$ planes, CuO chains, etc.) and have
treated the intra-layer scattering mechanisms approximately;
in particular, we have neglected the possibility of anisotropic
scattering, which may be important in $d$-wave superconductors.
Additionally, we have not attempted a fully self-consistent description
of the superconducting state, but have relied on BCS theory.
As we have argued, these assumptions are well justified.
We therefore expect that the results which follow will not be
strongly modified in more sophisticated treatments.

\subsection{Results}

{}From Eq.~(\ref{eq:j}), the total Josephson critical current is the
sum of three terms, each one corresponding to an inter-layer hopping
process.
Following the preceding discussion, we calculate in this
subsection the critical current from each inter-layer
process using the diagrams in Fig.~\ref{fig:diagrams} and
a BCS gap function.\cite{gapfn}

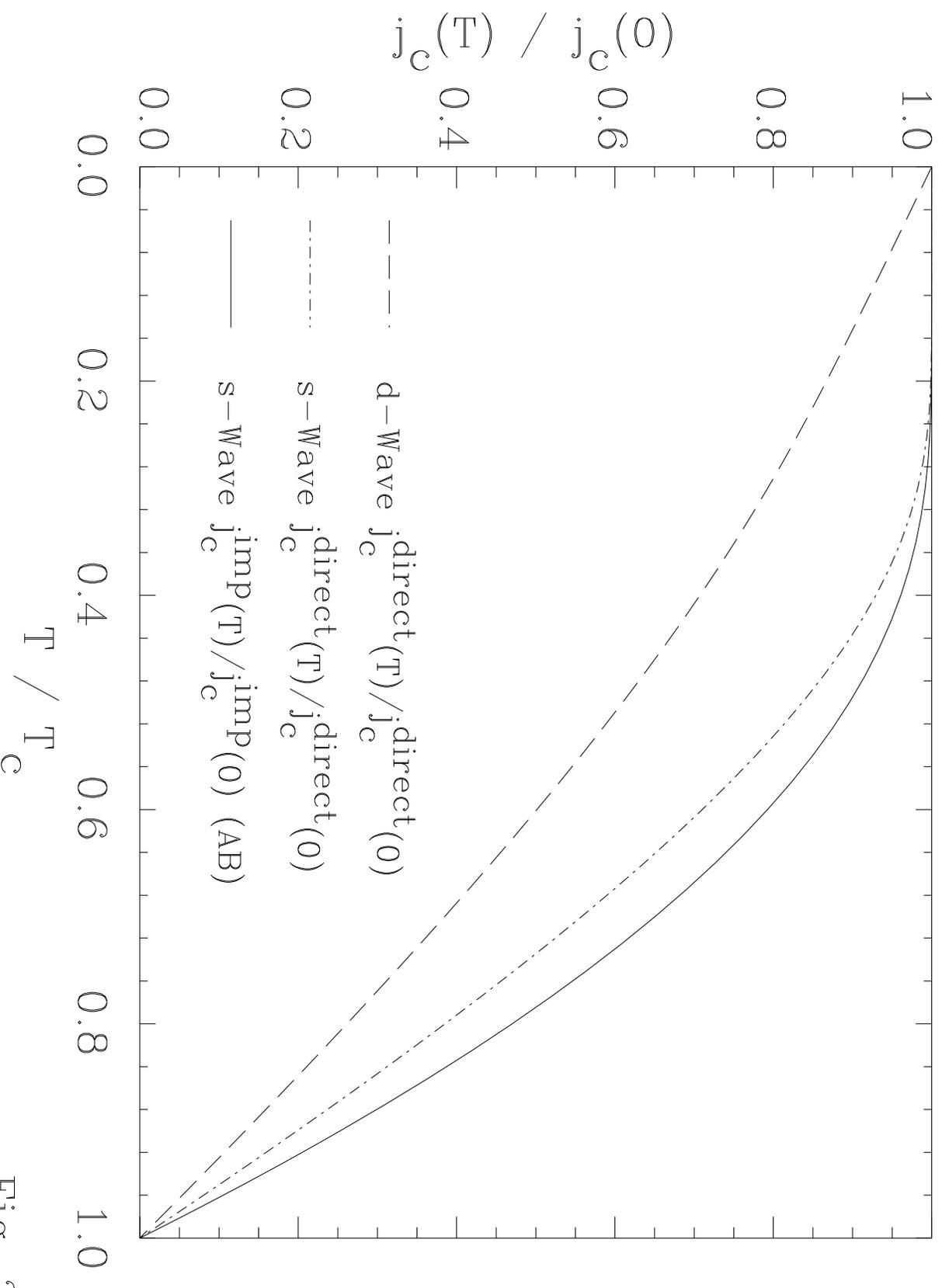
\begin{figure}
\caption{Normalized contributions to the $c$-axis critical current
$j_c (T) / j_c (0)$ as a function of the reduced temperature $T/T_c$
from various inter-layer hopping processes and pairing symmetries.
Shown are critical currents for $d$-wave pairing and direct hopping
($j_c^{\rm direct} (T) / j_c^{\rm direct} (0)$ from
Eq.~(\protect\ref{eq:j_direct}), dashed line),
$s$-wave pairing and direct hopping
($j_c^{\rm direct} (T) / j_c^{\rm direct} (0)$ from
Eq.~(\protect\ref{eq:j_direct}), dot-dashed line), and
$s$-wave pairing and impurity-assisted hopping
($j_c^{\rm imp} (T) / j_c^{\rm imp} (0)$ from
Eq.~(\protect\ref{eq:j_imp}), solid line).
These results are identical to, respectively, the in-plane BCS superfluid
density for $d$-wave (dashed line)
and $s$-wave pairing (dot-dashed line) and
the critical current in a macroscopic SIS tunnel junction
with $s$-wave pairing obtained within the Ambegaokar-Baratoff
(AB) theory\protect\cite{AB} (solid line).
Note that the $d$-wave impurity-assisted hopping critical current
vanishes for isotropic scattering.}
\label{fig:2Dc-2Dc}
\end{figure}

For direct hopping, this procedure gives
\begin{equation}
j_c^{\rm direct} (T) = \frac{2e}{\hbar a^2} \, N_0 t_\bot^2
   \left< \int_{-\omega_D}^{\omega_D} d\epsilon
   \frac{\Delta_{\bf k}^2}{E_{\bf k}^2}
   \, \left[
  \frac{\tanh \left( E_{\bf k} / 2 k_B T \right)}{2 E_{\bf k}} +
  \frac{\partial f (E_{\bf k})}{\partial E_{\bf k}}
  \right] \right>,
\label{eq:j_direct}
\end{equation}
where $E_{\bf k} = \sqrt{\epsilon^2 + \Delta_{\bf k}^2}$,
$f(x)$ is the Fermi distribution function, and the angle
brackets denote an average over the Fermi surface.\cite{error}
In the limit of zero-temperature,
$j_c^{\rm direct} (0) = (2e / \hbar a^2) \, N_0 t_\bot^2$ for either $s$- or
$d$-wave symmetry, {\it independently} of the magnitude of $\Delta_0 (0)$.
By contrast, $j_c (0) \propto \Delta_0 (0)$ in the Ambegaokar-Baratoff
formula.\cite{AB}
As a function of temperature,
$j_c^{\rm direct} (T) / j_c^{\rm direct} (0)$ varies exactly as the
in-plane BCS superfluid density, which is plotted in
Fig.~\ref{fig:2Dc-2Dc} for both $s$- and $d$-wave pairing.
At first sight, it is surprising that an incoherent treatment of
the direct interlayer hopping yields the same temperature dependence
as the (coherent) in-plane transport.
However, as in the normal state, the reason behind this behavior
derives from the fact that the vertices in the tunneling diagram
are diagonal in 2D wave vector and so give an effectively 3D result
(cf. the discussion after Eq.~(\ref{eq:o_direct})).
We also point out that $j_c^{\rm direct} (T)$ depends strongly on
the pairing
symmetry at low $T$, being nearly independent of $T$ for an isotropic
$s$-wave gap but proportional to $T$ for a clean $d$-wave gap, as might
be expected from the presence of nodes in $\Delta_{\bf k}$.

Impurity-assisted inter-layer hopping gives rise to a contribution
to the critical current which has the Ambegaokar-Baratoff form
if the order parameter is $s$-wave,\cite{AB}
\begin{equation}
j_c^{\rm imp} (T) = \frac{2e}{\hbar a^2} \, (\pi N_0 V)^2 \,
 \Delta_0(T) \tanh \left(\frac{\Delta_0(T)}{2k_BT}\right)
\label{eq:j_imp}
\end{equation}
[cf. the solid line in Fig.~\ref{fig:2Dc-2Dc}],
and vanishes by symmetry if the order parameter is $d$-wave
and the impurity scattering is isotropic.
The assumption of isotropic scattering is crucial if
$j_c^{\rm imp} (T)$ is to vanish for $d$-wave superconductors;
if the scattering is anisotropic, then the critical current is in
general finite.\cite{Graf2}
One generally expects impurity scattering to be mainly isotropic,
so $j_c^{\rm imp} (T)$ is probably small in $d$-wave
superconductors, but detailed calculations of the scattering
matrix element in these systems are required to make a
quantitative estimate.

Finally, the boson-assisted hopping processes also contribute to
the Josephson critical current.
Direct calculation of the relevant diagram yields
\begin{equation}
j_c^{\rm inel} = \frac{\pi e}{\hbar a^2} \, N_0 \lambda \, \,
 \Delta_0^2(0) \, \, I_{\rm inel} (T),
\label{eq:j_ba}
\end{equation}
where ($\hbar = k_B = 1$)
\begin{eqnarray}
I_{\rm inel} (T) &=& \frac{2}{\pi} \, \frac{\Delta_0^2(T)}{\Delta_0^2(0)} \,
 \int_{0}^{\omega_D} \frac{d\epsilon}{E} \,
 \int_{0}^{\omega_D} \frac{d\epsilon'}{E'} \nonumber \\
 &&\times \left\{
 \left[ f(-E') - f(E) \right]
 \frac{2\Omega_0}{(E' + E)^2 - \Omega_0^2}
 \left[ (E' + E) \coth \left( \frac{\Omega_0}{2T} \right)
 - \Omega_0 \coth \left( \frac{E' + E}{2T} \right) \right] \right. \nonumber \\
 &&+ \left. \left[ f(E') - f(E) \right]
 \frac{2\Omega_0}{(E' - E)^2 - \Omega_0^2}
 \left[ (E' - E) \coth \left( \frac{\Omega_0}{2T} \right)
 - \Omega_0 \coth \left( \frac{E' - E}{2T} \right) \right] \right\},
\label{eq:I_ba}
\end{eqnarray}
$\Omega_0$ is the Einstein phonon frequency,
$\lambda = 2 N_0 g^2 / \Omega_0$, and $s$-wave pairing is assumed.
For $d$-wave pairing and structureless electron-boson coupling,
the boson-assisted Josephson current vanishes.\cite{dsense}

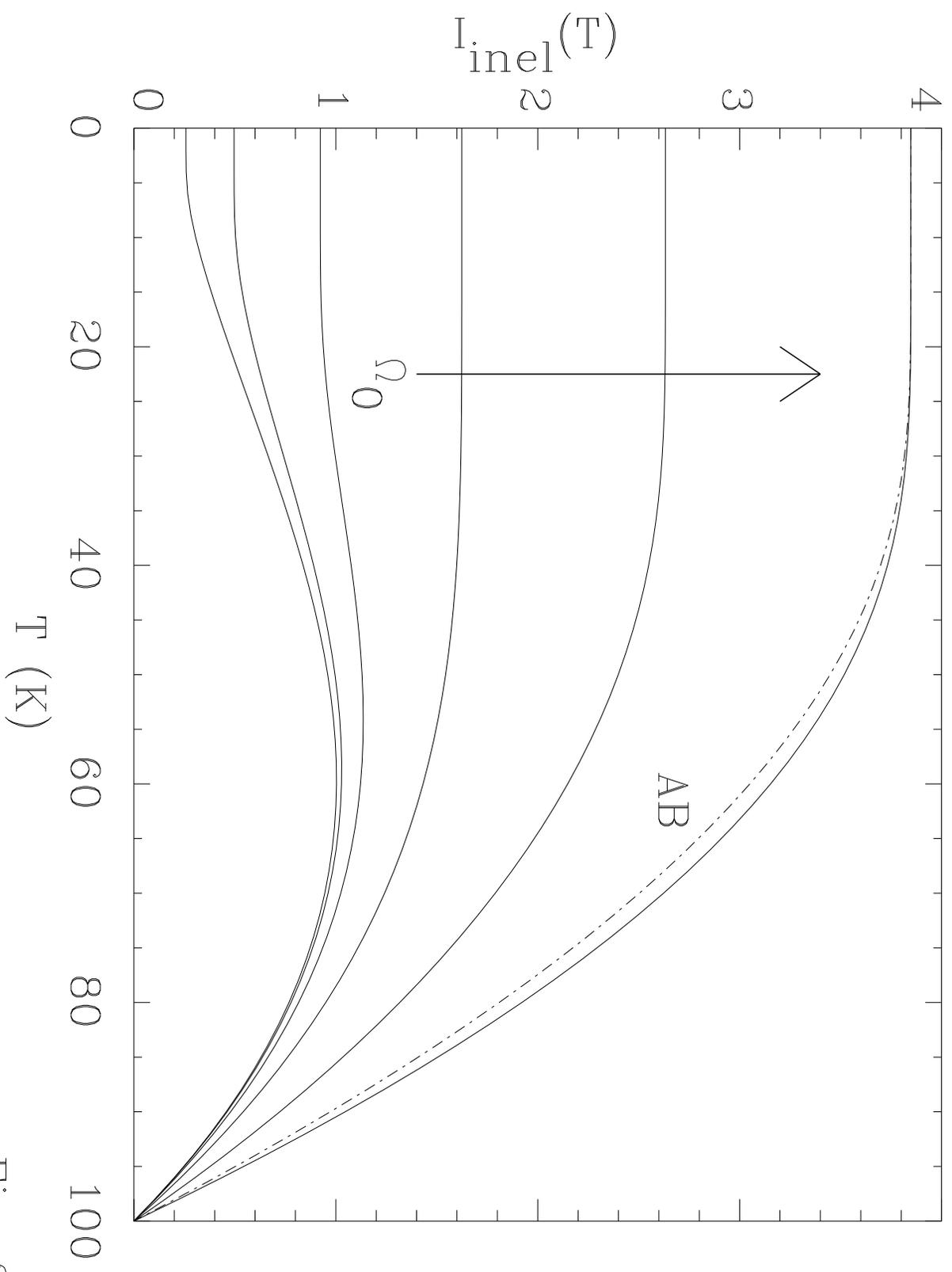
\begin{figure}
\caption{Normalized boson-assisted $c$-axis critical current
$I_{\rm inel} (T)$ [Eq.~(\protect\ref{eq:I_ba})]
as a function of temperature $T$ for Einstein
phonons with frequencies $\hbar \Omega_0 / k_B$ (from top to bottom) of
800 K, 400 K, 200 K, 100 K, 50 K, and 25 K with fixed
coupling constant $\lambda$ (solid lines; see text for details).
The temperature dependence of the gap function used in this
calculation is determined from the $s$-wave BCS equation with a
Debye temperature of 1000 K and a coupling constant fixed so
that the critical temperature is 100 K.
These choices yield $\Delta_0 (0)$ = 178 K.
For comparison, the Ambegaokar-Baratoff (AB)
result for an SIS Josephson junction\protect\cite{AB} normalized to
the largest $I_{\rm inel} (0)$ is also shown (dot-dashed line).}
\label{fig:Iinel}
\end{figure}

The $T$ and $\Omega_0$ dependence of $I_{\rm inel} (T)$ are
illustrated in Fig.~\ref{fig:Iinel} for a 100 K superconductor.
This figure shows several interesting features of $I_{\rm inel} (T)$.
First, we see that this function is approximately a non-zero constant at
low $T$, and the magnitude of this constant grows with $\Omega_0$.
Second, $I_{\rm inel} (T)$ resembles the
Ambegaokar-Baratoff result for $\Omega_0 > \Delta_0 (0)$,
but exhibits a peak when $\Omega_0 < \Delta_0 (0)$.
Finally, near $T_c$, $I_{\rm inel} (T)$ shows a linear
dependence on $T_c - T$.

We may explain these features simply.
First, the Josephson current does not vanish at zero
temperature because the zero-point energy of the bosons allows
the creation of virtual bosons to mediate the inter-layer
Cooper pair tunneling.
Since the zero-point energy is proportional to $\Omega_0$, the
increase of $I_{\rm inel} (0)$ with $\Omega_0$ is expected.
Second, when $\Omega_0 > \Delta_0 (0)$, $\Delta_0 (0)$
is the lowest energy
scale and controls the integral, so $I_{\rm inel} (T)$ is structureless.
On the other hand, when $\Omega_0 < \Delta_0 (0)$, we see from
Eq.~(\ref{eq:I_ba}) that the $\Omega_0 \coth (\Omega_0 / 2T)$
terms dominate the integral and lead to an
increase in $I_{\rm inel} (T)$ with increasing $T$.
At larger T, the decrease in $\Delta_0 (T)$ compensates for this rise
and eventually brings $I_{\rm inel} (T)$ to zero at $T_c$.
Third, the behavior near $T_c$ results from the fact that
$j_c^{\rm inel} (T) \propto \Delta_0^2 (T)$ as $\Delta_0(T)$ goes
to zero.

The different forms for the critical currents discussed in this
Section give rise to the possibility that measurements of $j_c (T)$
or the associated penetration depth may shed some light on the
order parameter symmetry and the relative importance of the
inter-layer hopping mechanisms in the cuprates.
We have demonstrated that the effects of pairing symmetry in this
simple model are quite strong at low temperatures:
$j_c (T) \propto T$ at low $T$ if the pairing is $d$-wave,
and $j_c (T)$ is exponentially close to $j_c (0)$ if it is isotropic
$s$-wave.
This conclusion follows from the fact that
$j_c^{\rm direct} (T) \propto T$
and $j_c^{\rm imp} (T) = j_c^{\rm inel} (T) = 0$ for $d$-wave
pairing, while all contributions are exponentially close to their
$T = 0$ values for isotropic $s$-wave pairing.
Even in a more sophisticated model that includes anisotropy in the
impurity- and boson-assisted scattering processes, the power-law
dependence of $j_c (T)$ at low $T$ for $d$-wave pairing should
persist due to the presence of the nodes in the gap function.
Consequently, the low-temperature behavior of the $c$-axis critical
current may be an important probe of the order parameter symmetry.
In addition, we have seen that each inter-layer hopping
process yields a $j_c$ with a distinct temperature dependence.
Looking at the temperature dependence of the total $j_c (T)$ may
therefore provide some clues about the relative importance
of the different inter-layer hopping processes.
The goal of the next section is to make a quantitative estimate
of the critical currents for optimally doped and de-oxygenated
YBCO and see if these expectations are realistic.

\section{Relation to Experiment}

The model we have presented builds on a natural connection between
the normal- and superconducting-state properties.
In this Section, we further exploit this connection by using experimental
results on the normal-state properties of
optimally doped and de-oxygenated YBCO to estimate $j_c (T)$
for these compounds.
In the next two subsections, we describe the method for
estimating the parameters in our model and extract the
zero-temperature critical currents and penetration depths.
We will use straightforward back-of-the-envelope reasoning
in making these estimates, but the magnitudes of these estimates
are confirmed by more detailed fits to normal-state transport
in this model.\cite{Lau,PhD}
In the succeeding subsection, we discuss the temperature
dependence of the resulting critical currents and determine
whether questions relating to the order parameter symmetry and
the inter-layer hopping mechanism can be addressed in measurements
of $j_c (T)$.
Finally, we interpret recent experimental measurements
based on the results of our calculations.

\subsection{Optimally Doped YBCO}

Experimentally, the resistivity in optimally doped ($T_c$ = 90 K)
YBCO is roughly
linear in temperature in both the in-plane and $c$-axis directions.
Within the incoherent hopping model, we can attribute this behavior
to purely direct hopping between the layers:
from Eq.~(\ref{eq:o_direct}), the intra-layer scattering rate is directly
reflected in the $c$-axis resistivity, implying that
$\rho_{ab} \propto \rho_c$.
In the superconducting state, this situation corresponds to
$j_c (T) \cong j_c^{\rm direct} (T)$.
To estimate the magnitude of this critical current, we observe that
the band structure $c$-axis plasma frequency
\begin{equation}
(\hbar \omega_c)^2 = \frac{4\pi e^2}{a^2 c} \sum_{{\bf k}\sigma}
 \delta(\epsilon_{\bf k}) \,
 \left( \frac{\partial \epsilon_{\bf k}}{\partial k_z} \right)^2 \cong
 16 \pi e^2 \frac{c}{a^2} N_0 t_{\bot}^2.
\end{equation}
The $T$ = 0 limit of Eq.~(\ref{eq:j_direct}) then yields
\begin{equation}
j_c^{\rm direct} (0) \cong \frac{(\hbar \omega_c)^2}{8\pi e \hbar c}.
\end{equation}
Optical measurements give $\hbar \omega_c$ = 400 meV,\cite{omegac}
from which we deduce
$j_c^{\rm direct} (0) \cong 9 \times 10^7$ A cm$^{-2}$.

In layered materials, this critical current would also be related to
the $c$-axis penetration depth measured in microwave experiments:
$\lambda_c = \sqrt{ \hbar c_l^2 / 8 \pi e c j_c^{\rm direct}}$,
where $c_l$ is the speed of light.\cite{Clem}
This relation gives $\lambda_c \cong$ 0.5 $\mu$m for optimally doped YBCO,
which is in reasonable accord with independent measurements of this
quantity.\cite{Basov}
In addition, Graf {\it et al.} estimate $\lambda_c \cong$
0.9 $\mu$m for optimally doped YBCO, purely disorder-assisted hopping,
and $s$-wave pairing,\cite{Graf} which is also in good agreement with our
(independent) result.
Assuming a density of states of 3 eV$^{-1}$,\cite{Pickett}
we estimate $t_{\bot} \approx$ 10 meV.
These results are summarized in the first column of
Table~\ref{Tab:Josephson}.

\subsection{De-Oxygenated YBCO}

To estimate the model parameters in the de-oxygenated case
is more complicated, since we expect all inter-layer hopping
processes to contribute to the critical current approximately equally.
In particular, the boson-assisted hopping contribution should be
significant in order to account for the upturn in $\rho_c$
at low temperatures.
We begin by examining experimental measurements of the $c$-axis
conductivity in de-oxygenated samples with $T_c$ = 40~K\cite{Veal}
and find that $\sigma_c \approx$ 1 $\Omega^{-1}$ cm$^{-1}$ near $T_c$
and $\sigma_c$ is approximately
linear in $T$ at high temperatures with a slope
$(d \sigma_c / dT)_{\rm high T} \approx 0.018 \,\, \Omega^{-1}$
cm$^{-1}$ K$^{-1}$.
These asymptotic results are expected within our model, as can be
seen from Eqs.~(\ref{eq:o_direct})-(\ref{eq:o_ba}) (and see below).

It is reasonable to assume (and detailed fits
demonstrate\cite{Lau,PhD}) that the boson-assisted component of the
normal-state conductivity has frozen out for $T \sim T_c$,
so we partition the remaining low-temperature
conductivity equally between the direct and
disorder-assisted processes.
In addition, the lattice parameters for YBCO\cite{Pickett} give
$\sigma_0 = $ 0.24 $\mu \Omega^{-1}$ cm$^{-1}$ in
Eqs.~(\ref{eq:o_direct})-(\ref{eq:o_ba}).
Combining these facts with Eq.~(\ref{eq:o_imp}) immediately
gives $N_0 V \approx 1.4 \times 10^{-3}$ and
$j_c^{\rm imp} (0) = 3.7 \times 10^4$ A cm$^{-2}$.
To extract the magnitude of the direct hopping, we note that
optical measurements give
$\hbar / \tau_{ab} = 2 \pi \lambda_{ab} k_B T$ with
$\lambda_{ab} =$ 0.2 to 0.4.\cite{Orenstein}
Taking $\lambda_{ab} =$ 0.3, one obtains $\hbar / \tau_{ab} \cong$
6 meV at $T_c$ = 40~K.
Inserting this result into Eq.~(\ref{eq:o_direct}) produces
$N_0 t_{\bot}^2 \approx 4 \times 10^{-5}$ meV or $t_{\bot} \approx$
0.1 meV if $N_0 = 3$ eV$^{-1}$.\cite{Pickett}
{}From $N_0 t_{\bot}^2$ and $\sigma_0$, we obtain
$j_c^{\rm direct} (0) \cong 1.4 \times 10^4$ A cm$^{-2}$.

At high temperatures, Eqs.~(\ref{eq:o_direct})-(\ref{eq:o_ba})
can be written
\begin{equation}
\sigma_c \sim \sigma_0 \left(
  \frac{N_0 t_{\bot}^2}{2 \pi^2 \lambda_{ab} k_B T}
  + (N_0 V)^2 + \lambda N_0 k_B T \right),
\end{equation}
from which we obtain
\begin{equation}
j_c^{\rm inel} (T) \cong \frac{1}{4ec} \,
 \left. \frac{d \sigma_c}{dT} \right|_{\rm high T} \, \Delta^2(0) \,
 I_{\rm inel} (T).
\end{equation}
To compute this term, we also need the optical
phonon frequency $\Omega_0$.
Recent $c$-axis-polarized Raman experiments indicate that the
500 cm$^{-1}$ (720 K) O(4) phonon is important in inter-layer
hopping,\cite{Nyhus} and so we take $\hbar \Omega_0 / k_B$ = 720 K.
Since $\hbar \Omega_0 >> \Delta_0 (0)$, Fig.~\ref{fig:Iinel} indicates
that $j_c^{\rm inel} (T)$ will monotonically decrease with $T$.
Numerical computation of $I_{\rm inel} (0)$ then gives
$j_c^{\rm inel} (0) = 12.5 \times 10^4$ A cm$^{-2}$.

The estimated parameters for de-oxygenated YBCO are summarized in
the second column of Table~\ref{Tab:Josephson}.
Observe that $t_{\bot}$ has been reduced by two orders
of magnitude from its fully oxygenated value:  $t_\bot \approx$
10 meV in $T_c =$ 90 K compounds, but $t_{\bot} \approx$
0.1 meV in $T_c =$ 40 K compounds.
This strong variation in $t_\bot$ with doping is
consistent with an exponentially decreasing wave-function overlap
induced by the increasing $c$-axis unit cell dimension\cite{Veal} and
by disorder and vacancies in the chains.
Empirically, the $c$-axis resistivity increases by
two orders of magnitude from optimally doped to de-oxygenated
samples,\cite{Cooper,Veal,Nyhus}
which would also naively correspond to at least an order-of-magnitude
decrease in $t_\bot$.
The data in Table~\ref{Tab:Josephson} also indicate that
$j_c^{\rm inel} (0):j_c^{\rm imp} (0):j_c^{\rm direct} (0) \cong$
9:2.6:1, indicating that the Josephson current
is dominated by the assisted processes in the de-oxygenated samples.
This result is quite reasonable, given the importance of this
term in determining the semiconducting temperature dependence of
$\rho_c (T)$.
As a final check to the consistency of our theory,
we note that $t_{\bot} \sim T_c$ in optimally doped compounds, indicating
weak incoherence, but $t_{\bot} << T_c$ in the de-oxygenated compounds,
which puts these compounds firmly in the incoherent regime.

\subsection{Estimated $j_c (T)$}

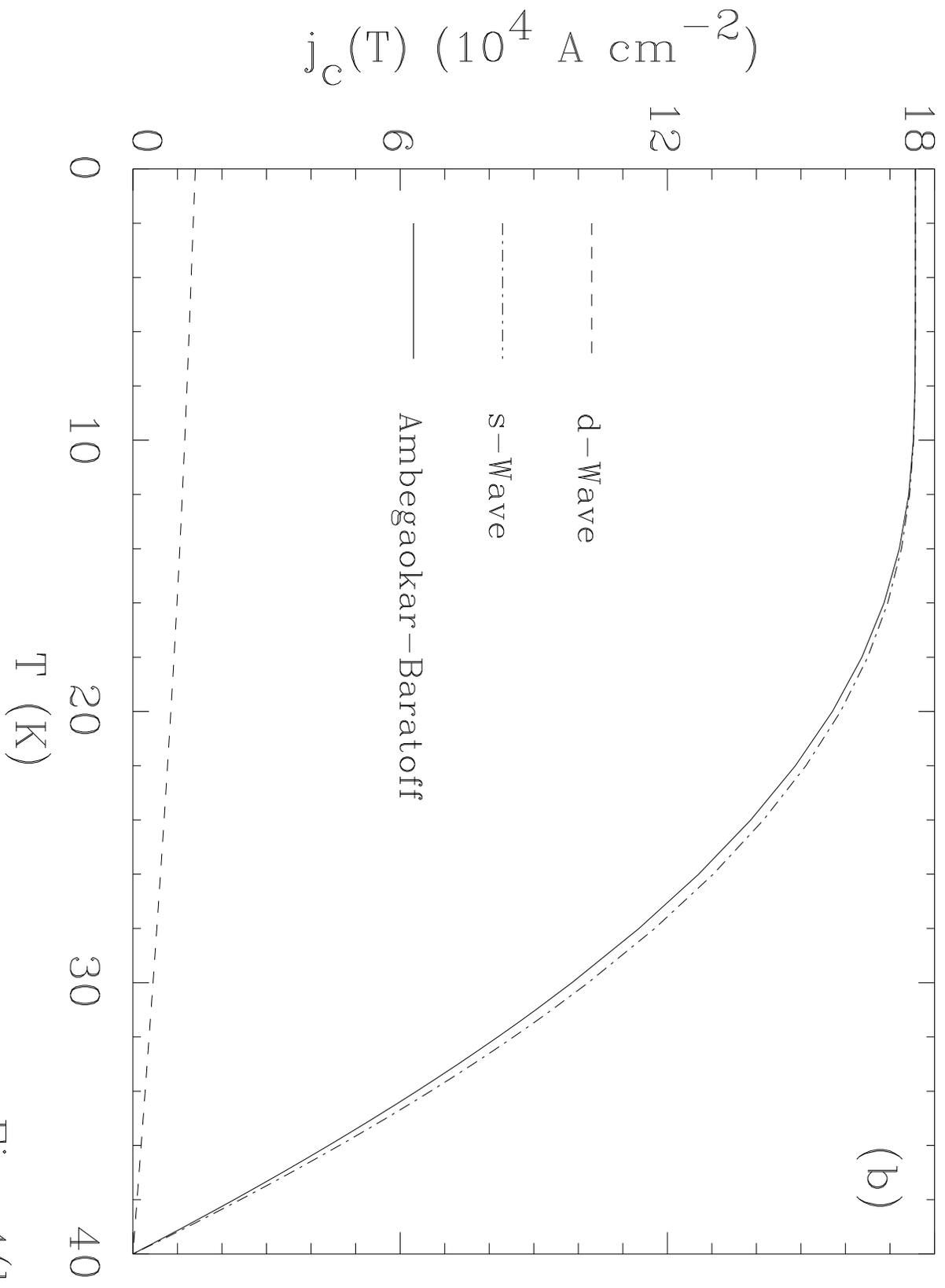
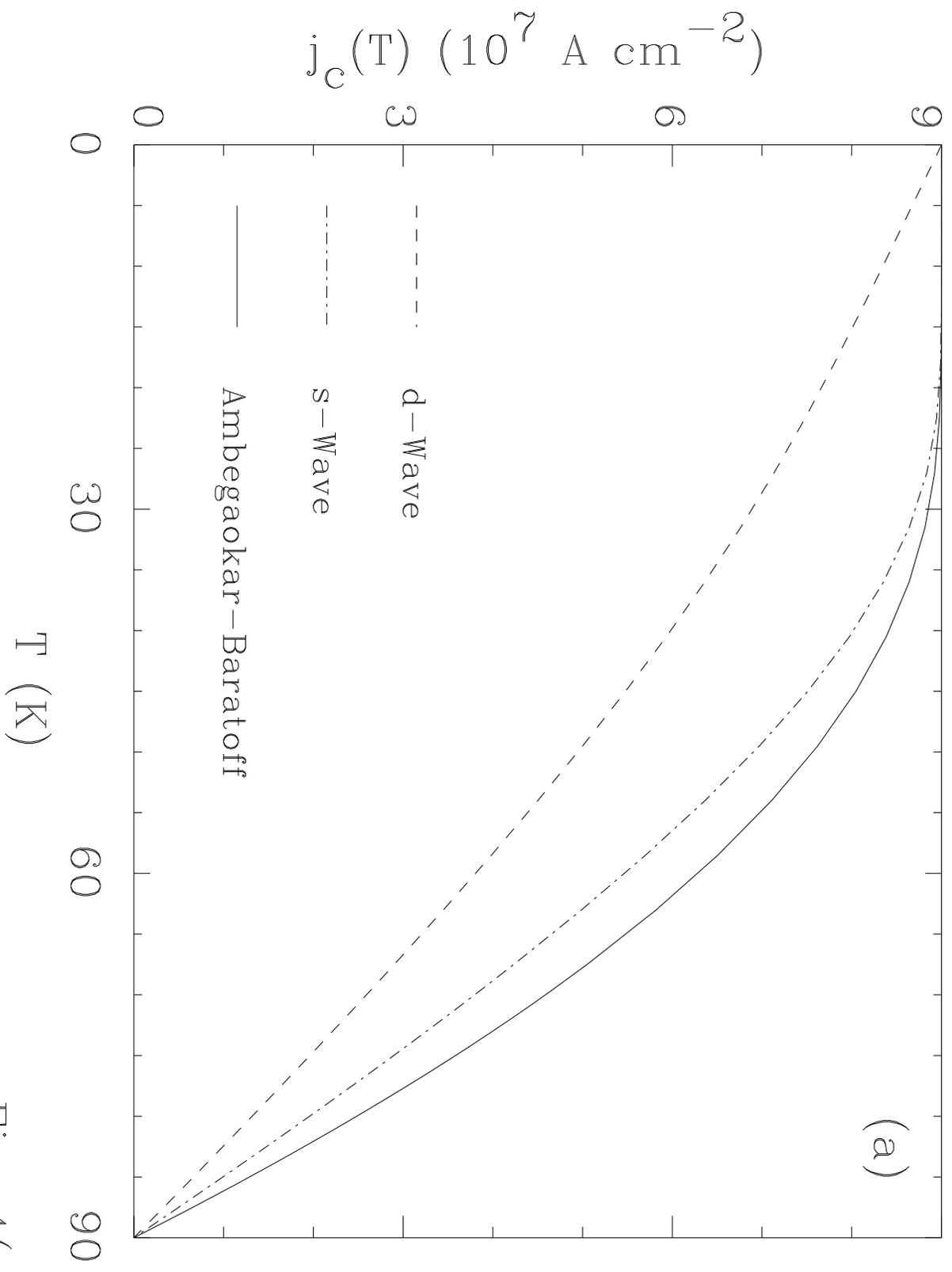
\begin{figure}
\caption{Estimated $c$-axis Josephson critical current $j_c (T)$
obtained within the incoherent hopping model as a
function of temperature $T$ for (a) optimally doped ($T_c =$ 90~K)
and (b) de-oxygenated ($T_c =$ 40~K) $\rm YBa_2Cu_3O_{7-\delta}$ with
$s$-wave (dot-dashed line) and $d$-wave (dashed line) pairing.
For comparison, the Ambegaokar-Baratoff result\protect\cite{AB}
normalized to the largest $j_c (0)$ is shown as a solid line.
The parameters used to generate these curves are discussed in
the text.}
\label{fig:pred}
\end{figure}

Using the parameters derived in the preceding two
subsections in Eqs.~(\ref{eq:j})-(\ref{eq:j_ba}), we can
compute the total $j_c (T)$ in both optimally doped and
de-oxygenated YBCO for both $s$- and $d$-wave paring.
The results of this calculation are shown in Fig.~\ref{fig:pred}.
For the optimally doped case [Fig.~\ref{fig:pred}(a)], we see
that the low-temperature $j_c (T)$ for $d$-wave
pairing can be clearly distinguished from $j_c (T)$ for $s$-wave
pairing, although $j_c(0)$ is independent of the pairing symmetry.
In addition, the critical currents for either pairing symmetry
are distinct from the conventional Ambegaokar-Baratoff result,
and this difference should be observable.
Specifically, the direct hopping dominates $j_c (T)$ in the
fully oxygenated case, so the observed critical current should
have the same temperature dependence as the in-plane
superfluid density.
Experimentally, this means that $\lambda_{ab}^2 (0) / \lambda_{ab}^2 (T)$
should equal $\lambda_{c}^2 (0) / \lambda_{c}^2 (T)$ in fully
oxygenated YBCO.

For the de-oxygenated case [Fig.~\ref{fig:pred}(b)], we again
find a clear difference between the critical currents for
$s$- and $d$-wave pairing at low temperature (although
it is not obvious from the figure).
However, $j_c (0)$ now depends strongly on the pairing symmetry
since the assisted processes are significant and contribute to
the $s$-wave $j_c$ but not to the $d$-wave $j_c$.
Additionally, while $d$-wave pairing yields a $j_c (T)$ very
different from the Ambegaokar-Baratoff result, the $s$-wave
critical current deviates from this form only very slightly.
Thus, if the pairing is $s$-wave, it is unlikely that a
measurement of $j_c (T)$ will enable one to deduce the relative
importance of the inter-layer hopping mechanisms.

The analysis of these figures allows us to draw two general
conclusions.
First, the order parameter symmetry should be indicated by the
low-temperature behavior of the measured critical current regardless
of oxygen content, being a power-law for $d$-wave and exponentially
close to a constant for isotropic $s$-wave.
This result is robust and independent of the
parameters estimated in the preceding two subsections.
On the other hand, the reduction in magnitude of the critical current
from $s$- to $d$-wave pairing in de-oxygenated samples
depends sensitively on the model parameters,
so it is unlikely that the pairing symmetry
will be indicated by an experimental measurement of $j_c (0)$,
unless the observed magnitude is exceedingly small (indicating
$d$-wave pairing).
Second, the presence of incoherent hopping in $\lambda_c$
should be revealed by the observation of either a 3D BCS
character if the pairing is $d$-wave or a 3D BCS character
that crosses over as a function of hole doping to a 2D
Ambegaokar-Baratoff character if the pairing is $s$-wave.
It is unlikely, however, that these measurements will in isolation
allow one to disentangle the relative contributions of the
different inter-layer hopping processes in the de-oxygenated samples.

\subsection{Comparison to Experiment}

Our model can be compared to direct experimental measurements of
$j_c (T)$ or to indirect measurements of this quantity through, for
example, microwave penetration depth measurements.
In the latter experiments, the penetration depth $\lambda_c$ is
related to the critical current $j_c$ by the relation
$j_c (T) / j_c (0) = \lambda_c^2 (0) / \lambda_c^2 (T)$.\cite{Clem}
Given the results of the preceding subsection,
we should ideally compare our theoretical estimates to measurements
of the critical current or penetration depth in YBCO.
Unfortunately, the experimental situation is not yet resolved
in this cuprate:
some penetration depth measurements show
an SNS-like $\lambda_c^2 (0) / \lambda_c^2 (T)$ which is
incompatible with our model,\cite{Anlage}
while others yield more conventional curves.\cite{Bonn}
However, we may still examine measurements on other cuprates and
look for some of the generic features of our model.

The most thoroughly studied cuprate with regard to intrinsic
Josephson junction effects is $\rm Bi_2Sr_2CaCu_2O_8$
(BSCCO).\cite{KM}
Direct critical current measurements are available for this
material that show Ambegaokar-Baratoff behavior in
oxygen-annealed samples and a temperature-independent
$j_c(T)$ at low $T$ that falls rapidly to zero near
$T_c$ in argon-annealed samples.\cite{KM}
Within our model, the Ambegaokar-Baratoff-like $j_c (T)$ in the
oxygen-annealed samples suggests $s$-wave pairing and a
significant contribution to $j_c$ from assisted hopping, which is
analogous to de-oxygenated YBCO [cf. Fig.~\ref{fig:pred}(b)].
The approximately constant $j_c(T)$ in the argon-annealed
samples is also indicative of $s$-wave pairing, but the rapid
reduction of $j_c$ near $T_c$ is difficult to explain within our model.

In addition to BSCCO, recent measurements of the $c$-axis penetration
depth on $\rm La_{2-x}Sr_xCuO_4$ (LSCO) have been reported and
interpreted within an intrinsic Josephson effect picture.\cite{Shibauchi}
As with oxygen-annealed BSCCO, LSCO shows an approximately
Ambegaokar-Baratoff $j_c (T)$, which we interpret as evidence
for significant assisted inter-layer hopping as well as $s$-wave
pairing.
However, the data do not extend to very low temperatures, so
additional experimental measurements in this compound
(and in BSCCO) would be illuminating.

\section{Conclusions}

In this paper, we have considered an incoherent hopping model
that reproduces the features of the $c$-axis resistivity in both
optimally doped and underdoped cuprates.
We have assumed that there are three processes which transport
quasiparticles from one CuO$_2$ layer to another:
a direct hopping induced by wave-function overlap, an impurity-assisted
hopping due to disorder, and a
boson-assisted hopping induced by, for example, modulation of the
inter-layer distance by phonons.
By performing calculations to second order in the inter-layer
hopping amplitudes, we have shown that
this model yields an intrinsic Josephson effect in the superconducting
state, and we have analyzed the contributions to the
Josephson critical current arising from these three processes.
We find a robust dependence of the low-temperature $c$-axis
critical current on the order parameter symmetry which should
be reflected in experimental measurements of $j_c (T)$
or the $c$-axis microwave penetration depth ratio
$\lambda_c^2 (0) / \lambda_c^2 (T)$.

This conclusion is supported by the application of our model
to YBCO, where we have estimated the magnitude and temperature
dependence of $j_c$ and find a clear-cut signature of the pairing
symmetry in the low-temperature behavior of $j_c (T)$ for both
fully oxygenated and de-oxygenated compounds.
Additionally, our model predicts that the $c$-axis penetration
depth ratio should have either a 3D BCS character if the pairing is
$d$-wave or a 3D BCS character which crosses over as a function
of hole doping to a 2D Ambegaokar-Baratoff character if the
pairing is $s$-wave.
Measurements on BSCCO and LSCO suggest that the pairing
in these materials is $s$-wave and the inter-layer hopping
has a significant assisted component, but direct comparison of our
estimates with YBCO is not possible because the data have not
yet converged.
Further experimental work on both fully oxygenated and
de-oxygenated YBCO is clearly required and would provide a
stringent test of our theory.

\acknowledgements

The authors would like to thank B. W. Veal for providing us with the
resistivity data for YBCO and C. N. Lau for fitting
these data with the Rojo-Levin model.
We would also like to thank K. Beauchamp, K. E. Gray,
H. M. Jaeger, A. J. Leggett, M. Ledvij, A. G. Rojo,
L. Radzihovsky, T. F. Rosenbaum, T. Timusk, and N. Trivedi for
helpful conversations.
This work was supported by the National Science Foundation
(DMR 91-20000) through the Science and Technology Center for
Superconductivity (KL, RJR) and by DMR-MRL-8819860 and
DMR-9123577 (RJR).

\begin{table}
\caption{Estimates of the direct hopping amplitude $t_{\bot}$, the
zero-temperature critical currents, and the zero-temperature penetration
depths for fully oxygenated ($T_c$ = 90 K) and de-oxygenated ($T_c$ = 40 K)
$\rm YBa_2Cu_3O_{7 - \delta}$ within the incoherent hopping model.
The critical currents are presented for the direct
($j_c^{\rm direct} (0)$),
impurity-assisted ($j_c^{\rm imp} (0)$),
and boson-assisted ($j_c^{\rm inel} (0)$)
inter-layer hopping processes along with the total critical
currents ($j_c^s (0)$, $j_c^d (0)$) and corresponding penetration
depths ($\lambda_c^s (0)$, $\lambda_c^d (0)$)
for $s$- and $d$-wave pairing.
See text for details.}
\begin{tabular}{lcc}
& Optimally Doped & De-oxygenated \\
\tableline
$t_{\bot}$ (meV) & 10 & 0.1 \\
$j_c^{\rm direct} (0)$ (A cm$^{-2}$)
& 9 $\times$ 10$^7$ & 1.4 $\times$ 10$^4$ \\
$j_c^{\rm imp} (0)$ (A cm$^{-2}$) & $\sim$ 0 & 3.7 $\times$ 10$^4$ \\
$j_c^{\rm inel} (0)$ (A cm$^{-2}$) & $\sim$ 0 & 12.5 $\times$ 10$^4$ \\
 \\
$j_c^s (0)$ (A cm$^{-2}$)& 9 $\times$ 10$^7$ & 17.6 $\times$ 10$^4$ \\
$j_c^d (0)$ (A cm$^{-2}$)& 9 $\times$ 10$^7$ & 1.4 $\times$ 10$^4$ \\
 \\
$\lambda_c^s (0)$ ($\mu$m)& 0.5 & 11 \\
$\lambda_c^d (0)$ ($\mu$m)& 0.5 & 40 \\
\end{tabular}
\label{Tab:Josephson}
\end{table}


\begin{references}

\bibitem{Cooper}For a review, see S. L. Cooper and K. E. Gray, in
{\it Physical Properties of High-Temperature Superconductors IV},
D. M. Ginsberg, ed. (World Scientific, Singapore, 1994).

\bibitem{KumarL}N. Kumar, P. A. Lee, and B. Shapiro, Physica A {\bf
168}, 447 (1990).

\bibitem{Tachiki}S. Takahashi and M. Tachiki, Physica C {\bf 170},
505 (1990); M. Tachiki, T. Koyama, and S. Takahashi, Prog. Theor.
Phys. Suppl. {\bf 108}, 297 (1992).

\bibitem{Klemm}R. A. Klemm and S. H. Liu, Phys. Rev. Lett.
{\bf 74}, 2343 (1995); Phys. Rev. B {\bf 44},  7526 (1991);
and references therein.

\bibitem{KumarJ}N. Kumar and A. M. Jayannavar, Phys. Rev. B
{\bf 45}, 5001 (1992).

\bibitem{Leggett}A. J. Leggett, Braz. J. Phys. {\bf 22}, 129 (1992).

\bibitem{Graf}M. J. Graf, D. Rainer, and J. A. Sauls, Phys. Rev. B
{\bf 47}, 12089 (1993), and unpublished.

\bibitem{Rojo}A. G. Rojo and K. Levin, Phys. Rev. B {\bf 48}, 16861
(1993).

\bibitem{AZ}P. W. Anderson and Z. Zhou, Phys. Rev. Lett. {\bf 60},
132 (1988); {\it ibid.} {\bf 60}, 2557 (1988).

\bibitem{Chakravarty}S. Chakravarty, A. Sudb\o, P. W. Anderson,
and S. Strong, Science {\bf 261}, 337 (1993); A. Sudb\o, S. Chakravarty,
S. Strong, and P. W. Anderson, Phys. Rev. B {\bf 49}, 12245 (1994).

\bibitem{KM}R. Kleiner, F. Steinmeyer, G. Kunkel, and P. M\"{u}ller,
Phys. Rev. Lett. {\bf 68}, 2394 (1992); R. Kleiner and P. M\"{u}ller,
Phys. Rev. B {\bf 49}, 1327 (1994).

\bibitem{Cohen}M. H. Cohen, L. M. Falicov, and J. C. Phillips, Phys.
Rev. Lett. {\bf 8}, 316 (1962).

\bibitem{AB}V. Ambegaokar and A. Baratoff, Phys. Rev. Lett.
{\bf 10}, 486 (1963); {\it ibid.} {\bf 11}, 104 (1963).

\bibitem{Duke}C. B. Duke, {\it Tunneling in Solids} (Academic Press,
New York, 1969).

\bibitem{TMahan}G. D. Mahan, {\it Many-Particle Physics} (Plenum,
New York, 1986), Chapter 9.

\bibitem{Kleinman}L. Kleinman, B. N. Taylor, and E. Burstein,
Rev. Mod. Phys. {\bf 36}, 208 (1964); for a more recent review
concerned with device applications, see J. R. Tucker and M. J. Feldman,
Rev. Mod. Phys. {\bf 57}, 1055 (1985).

\bibitem{Kirtley}C. C. Tsuei, J. R. Kirtley, C. C. Chi,
L. S. Yu-Jahnes, A. Gupta, T. Shaw, J. Z. Sun, and M. B. Ketchen,
Phys. Rev. Lett. {\bf 73}, 593 (1994);
J. R. Kirtley, C. C. Tsuei, J. Z. Sun, C. C. Chi,
L. S. Yu-Jahnes, A. Gupta, M. Rupp, and M. B. Ketchen,
Nature {\bf 373}, 225 (1995).

\bibitem{Bulaevskii}L. N. Bulaevskii, V. V. Kuzii, and A. A. Sobyanin,
Solid State Commun. {\bf 25}, 1053 (1978).

\bibitem{Graf2}M. J. Graf, M. Palumbo, D. Rainer, and J. A. Sauls,
unpublished.

\bibitem{Clem}L. N. Bulaevskii, Sov. Phys.--JETP {\bf 37}, 1133 (1973);
J. R. Clem, Physica (Amsterdam) {\bf 162-164C}, 1137 (1989).

\bibitem{Lau}R. J. Radtke, C. N. Lau, and K. Levin, J. Supercon., August,
1995 (to appear).

\bibitem{Ioffe}L. B. Ioffe, A. I. Larkin, A. A. Varlamov, and L. Yu,
Phys. Rev. B {\bf 47}, 8936 (1993).

\bibitem{AM}For reviews, see J. P. Carbotte, Rev. Mod. Phys. {\bf 62},
1027 (1990), and P. B. Allen and B. Mitrovi\'{c}, in
{\it Solid State Physics}, H. Ehrenreich, F. Seitz, and D. Turnbull,
eds. (Academic Press, New York, 1982), Vol. 37, p. 1.

\bibitem{AGD}A. A. Abrikosov, L. P. Gor'kov, and I. E. Dzyaloshinski,
{\it Methods of Quantum Field Theory in Statistical Physics}
(Dover Publications, New York, 1963), pp. 325-341.

\bibitem{LV}P. B. Littlewood and C. Varma, Phys. Rev. B {\bf 45},
12636 (1992).

\bibitem{PhD}R. J. Radtke, Ph.D. Thesis, The University of Chicago,
Chicago, Illinois, 1994.

\bibitem{BB}In the strictest sense, Bloch-Boltzmann transport
is violated in this case as well, since the $c$-axis mean free path
is smaller than the $c$-axis unit cell distance.
Cf. Ref.~\protect\onlinecite{Leggett}.

\bibitem{fluc}Note that our theory does not account for fluctuation
effects near $T_c$, but these effects can be included if
necessary.
See Ref.~\protect\onlinecite{Ioffe} and
V. V. Dorin, R. A. Klemm, A. A. Varlamov, A. I.
Buzdin, and D. V. Livanov, Phys. Rev. B {\bf 48}, 12951 (1993).

\bibitem{Anderson}P. W. Anderson, J. Phys. Chem. Solids {\bf 11},
26 (1959).

\bibitem{Hirschfeld}L. S. Borkowski and P. J. Hirschfeld, Phys. Rev. B
{\bf 49}, 15404 (1994); R. Fehrenbacher and M. R. Norman,
Phys. Rev. B {\bf 50}, 3495 (1994).

\bibitem{gapfn}Specifically, we take $\Delta ({\bf k}, i\omega_n)$ =
$\Delta_{\bf k} \, \theta (\omega_D - |\omega_n|)$, where
$\Delta_{\bf k} = \Delta_0 (T) \, \eta_{\bf k}$ and
$\eta_{\bf k} = 1$ for $s$-wave and $\cos 2\phi$ for $d$-wave symmetry.
Here, $\phi$ is the angular coordinate along the 2D Fermi surface
(assumed circular), and the temperature dependence of $\Delta_0$ is
determined from the BCS equation.
In the numerical calculations, we take $\omega_D$ = 1000 K.

\bibitem{error}Note that Eq.~(\protect\ref{eq:j_direct}) contains
a term mistakenly omitted in Ref.~\protect\onlinecite{PhD}.

\bibitem{dsense}The disappearance of the $d$-wave $j_c^{\rm inel} (T)$
is a consequence of the simple model used and would
generally not apply if structure in the electron-boson coupling
$g$ or dispersion of the inter-layer bosons were included.
As with the impurity-assisted term, detailed calculations are
required in order to make a quantitative estimate of the residual
critical current in $d$-wave superconductors, but we expect that its
magnitude would be small.

\bibitem{omegac}S. L. Cooper, P. Nyhus, D. Reznik, M. V. Klein,
W. C. Lee, D. M. Ginsberg, B. W. Veal, A. P. Paulikas, and
B. Dabrowski, Phys. Rev. Lett. {\bf 70}, 1533 (1993).

\bibitem{Basov}D. N. Basov, T. Timusk, B. Dabrowski, and
J. D. Jorgensen, Phys. Rev. B {\bf 50}, 3511 (1994).

\bibitem{Pickett}W. E. Pickett, Rev. Mod. Phys. {\bf 61}, 433 (1989).

\bibitem{Veal}We have used data from B. W. Veal, A. P. Paulikas and
P. Kostic, unpublished, for these estimates, but other measurements
yield similar results; see Ref.~\protect\onlinecite{Cooper}.

\bibitem{Orenstein}J. Orenstein, G. A. Thomas, A. J. Millis, S. L. Cooper,
D. H. Rapkine, T. Timusk, L. F. Schneemeyer, and J. V. Waszczak,
Phys. Rev. B {\bf 42}, 6342 (1990).

\bibitem{Nyhus}P. Nyhus, M. A. Karlow, S. L. Cooper, B. W. Veal,
and A. P. Paulikas, Phys. Rev. B {\bf 50}, 13898 (1994).

\bibitem{Anlage}J. Mao, D. J. Wu, J. L. Peng, R. L. Greene, and
S. M. Anlage, Phys. Rev. B {\bf 51}, 3316 (1995).

\bibitem{Bonn}C. C. Homes, T. Timusk, D. A. Bonn, R. Liang, and
W. N. Hardy, submitted to Physica~C.

\bibitem{Shibauchi}T. Shibauchi, H. Kitano, K. Uchinokura, A. Maeda,
T. Kimura, and K. Kishio, Phys. Rev. Lett. {\bf 72}, 2263 (1994).

\end{references}
\end{document}